\documentclass[12pt]{article}
\include{epsf}
\include{psfig}
\include{epsfig}
\usepackage{epsfig}
\makeatletter

\def\compoundrel#1\over#2{\mathpalette\compoundreL{{#1}\over{#2}}}
\def\compoundreL#1#2{\compoundREL#1#2}
\def\compoundREL#1#2\over#3{\mathrel
         {\vcenter{\hbox{$\m@th\buildrel{#1#2}\over{#1#3}$}}}}
\makeatother

\makeatletter
\def\sss{\scriptscriptstyle}
\def\sVEV#1{\left\langle #1\right\rangle}
\def\nn{\hspace{2mm}}
\newcommand{\MeV}{\mbox{\rm MeV}}
\newcommand{\GeV}{\mbox{\rm GeV}}
\newcommand{\eV}{\mbox{\rm eV}}
\def\sleq{\raisebox{-.6ex}{${\textstyle\stackrel{<}{\sim}}$}}

\begin{document}
\setcounter{page}{0}
\thispagestyle{empty}
\setlength{\parindent}{1.0em}
\begin{flushright}
GUTPA/03/08/01
\end{flushright}
\renewcommand{\thefootnote}{\fnsymbol{footnote}}
\begin{center}{\LARGE{{\bf  Trying to understand the Standard Model 
parameters\footnote{\it To be published in the Proceedings of the XXXI ITEP 
Winter School of Physics, Moscow, Russia, 18 - 26 February 2003.} }}}
\end{center}
\begin{center}{\large{C. D. Froggatt}
%~\footnote[2]{E-mail: c.froggatt@physics.gla.ac.uk}\\}
\\}
\end{center}
\renewcommand{\thefootnote}{\arabic{footnote}}
\begin{center}{{\it Department of Physics and Astronomy}\\{\it
University of Glasgow, Glasgow G12 8QQ, Scotland}}\end{center}
\begin{center}{\large{H. B.
Nielsen}
%~\footnote[1]{E-mail: hbech@iph.alf.nbi.dk}\\}
\\}
\end{center}
\renewcommand{\thefootnote}{\arabic{footnote}}
\begin{center}{{\it  Niels Bohr Institute, }\\{\it
Blegdamsvej 17-21, DK 2100 Copenhagen, Denmark}}\end{center}
\setcounter{footnote}{0}

\begin{abstract}
We stress the importance of the circa 20 parameters in the
Standard Model, which are not fixed by the model but only
determined experimentally, as a window to the physics beyond the
Standard Model. However, it is a tiny window in as far as these
numbers contain only the information corresponding to about one
line of text. Looking for a method to study these coupling and
mass parameters, we put forward the idea of the Multiple Point
Principle as a first step. This principle states that Nature 
adjusts the coupling and mass parameters so as to make many different 
vacuum states exist and have approximately the same energy densities
(cosmological constants). As an illustrative application, we put
up the proposal that a small increase (maybe only an infinitesimal
one) in the value of the top quark coupling constant could lead to
a new vacuum phase; in this new phase the binding of a bound state
of 6 top quarks and 6 anti-top quarks becomes so strong as to
become a tachyon and condense in the vacuum. Assuming the
existence of a third degenerate vacuum at the fundamental energy
scale, we present a solution to the hierarchy problem of why the
ratio of the fundamental scale to the electroweak scale is so
large. We also present a 5 parameter fit to the orders of
magnitude of the quark-lepton masses and mixing angles in the
Family Replicated Gauge Group Model. In this model, the Standard
Model gauge group and a gauged B-L (baryon number minus lepton
number) is extended to one set of gauge fields for each family of
fermions.
\end{abstract}
%%%%%%%%%%%%%%%%%%%%%%%%%%%%%%%%%%%%%%%%%%%%%%%%%%%%%%%
\thispagestyle{empty}

\newpage

\section{Introduction}
\label{introduction}

A major challenge for physics today is to find the fundamental
theory beyond the Standard Model (the ``Theory of Everything'').
However, we have the difficulty that the vast majority of the
available experimental information is already, at least in
principle\footnote{One believes that many of the unexplained
details are due to the difficulties in making strong coupling QCD
calculations.}, explained by the Standard Model (SM). Also, until now,
there has been no convincing evidence for the existence of any
particles other than those of the SM and states composed of SM
particles. All accelerator physics seems to fit well with the SM,
except for neutrino oscillations. Apart from the neutrino masses
and mixing angles, the only phenomenological evidence for going
beyond the SM comes from cosmology and astrophysics. It is well-known
that the pure SM predicts a too low value for the baryon number
resulting from the big bang. In fact sphaleron transitions would
tend to wash out a possibly pre-existing excess. It is only
an excess of the (B-L) number which is conserved in the SM
that, if present at the electroweak era, could  produce the observed,
or rather fitted, baryon number in the early universe. Such a
(B-L) excess could naturally come out of a see-saw neutrino
model, which of course is also suitable for generating neutrino
masses and mixings. Other extensions of the SM, such as the
introduction of supersymmetric particles at the electroweak scale,
might also be able to prevent the wash-out of the baryon number
asymmetry. In addition to the baryon asymmetry, an explanation
for the dark matter in the universe, or some modification of the
gravitation from which it is deduced to exist, is needed.

Apart from these astrophysical problems, there is only very weak
experimental evidence for effects which do not match the SM
extended to include neutrino masses. This means that we have very
little knowledge about the true model beyond the SM, except for the
circa 20 parameters of the SM, neutrino oscillation data and
astrophysical-cosmological phenomena. Even the information
from astrophysics can only provide rather few parameters
which can be used to tell us something about the
theory beyond the SM. For dark matter, we mainly know the total amount
and indications of its clustering may give a hint to its interactions.
However we know little about the mass of dark matter particles (if
indeed dark matter does consist of particles). So, like the baryon
asymmetry in the universe, dark matter provides us with essentially
one single useful number. With the detailed studies of the microwave
background radiation, the inflationary period in the early universe
has been made accessible to phenomenological investigation. Thus there
are here also, say, 2 experimentally accessible parameters that could be
useful in finding the theory beyond the SM. In addition the
neutrino oscillation data provide us with two mass squared differences
and two mixing angles\footnote{Here
we ignore the possible LSND effect and sterile neutrinos and
treat the CHOOZ measurement as just an upper bound on the
$\theta_{e3}$ mixing angle.}. Thus we may claim to have about 8
parameters from physics beyond the SM: 4 parameters from cosmology
and 4 parameters from neutrino oscillations. One might consider
adding the cosmological constant $\Lambda_{cosmo}$ and Newton's
gravitational constant $G$ to these. However, even then, we only
obtain about 10 beyond the SM parameters compared to the circa 20
parameters in the SM Lagrangian. So our hope of getting numerical
checks, in addition to bounds, for a candidate beyond the SM theory
consists in fitting a total of about 30 parameters.

It is possible \cite{RughHolger} to estimate the amount of information
contained in our present knowledge of the SM parameters as well as
in the structure of the SM. This SM structure refers to the SM gauge
group and the representations of the latter under which the SM
particles transform. It turns out that this information makes
up about 200 bits, which means hardly one line of text. Even adding say
5 bits of information for knowing the order of magnitude of the baryon
asymmetry, 3 bits for the dark matter, 7 bits for the cosmological
constant and 12 bits from inflation, we still barely reach one line of
text, which may be counted as $5 \times 70 \simeq 350$ bits. We see that
the major part of the information\footnote{The remaining circa $20\%$
of the bit content is provided by the astrophysical parameters.} is
contained in the circa 20 parameters of the SM Lagrangian together with
the 4 neutrino mass and mixing parameters.

It is of course crucial for trusting a new theory that it should
have some support from experiment. So an important test for
any candidate model for physics beyond the SM physics is whether or
not it can predict or significantly fit the above mentioned SM
coupling constant and mass parameters. A simple quantum field theory
will only be able to do so by introducing a supplementary symmetry
or some other restricting principle.

String theory should, in principle, be able to do the job, in as far
as it has zero dimensionless parameters. However, in order to  avoid
immediate disagreement with, for example, the number of space-time
dimensions, one has to assume in practice the existence of a rather
complicated background space, which can easily introduce several
hard to calculate parameters. It is also, in practice, hard to
invent convincing specific supersymmetry breaking schemes. So the
predictions from string theory about the SM couplings and masses
inevitably end up depending on ad hoc assumptions and parameterisations.

In the present talk we shall consider quantum field theory models
containing the SM as a low energy limit or even simply the SM itself.
Our approach to predicting SM parameters is to impose some physically
observable (i.e.~cut-off independent) and relatively simple properties.
Symmetry relations provide an example of such restrictions on the
theory. A priori one could specify the energy, say, of a certain
configuration of particles or a certain S-matrix element. However it
can easily turn out that the number of kinematic specifications 
required to define the quantity under consideration 
is large and then it is not so obvious how to make these specifications
in a {\em simple} way. Therefore one would think that the simplest choice
would be to impose our restrictions on situations with as few particles
as possible. This remark suggests that the most promising approach is
to impose restrictions on the zero-particle state(s)---the vacuum or
vacua.

Indeed we already have to restrict the vacuum due to the
cosmological constant problem. We need to postulate that the
energy density in the vacuum in which we live is practically zero
relative to the Planck scale or even the electroweak scale. A
priori it is quite possible for a quantum field theory to have
several minima of its effective potential as a function of its
scalar fields, i.e.~there could be several vacua. So the question
now arises concerning the postulate of zero cosmological constant:
should the energy density, i.e.~the cosmological constant, be
(approximately) zero for all possible vacua or should it only be
zero for that vacuum in which we live? Of course it is only
phenomenologically required to assume that the cosmological
constant be zero in the vacuum in which we live. However the
assumption would {\em not} be {\em more complicated}, if we
postulated that all the vacua which might exist, as minima of the
effective potential, should have approximately zero cosmological
constant. The extension of the zero cosmological constant
assumption in this way corresponds to what we call the Multiple
Point Principle (MPP), which may be stated in the following form:
there are many ``vacua" with approximately the same energy density
or cosmological constant.

\begin{figure}[t]
\leavevmode
\centerline{\epsfig{file=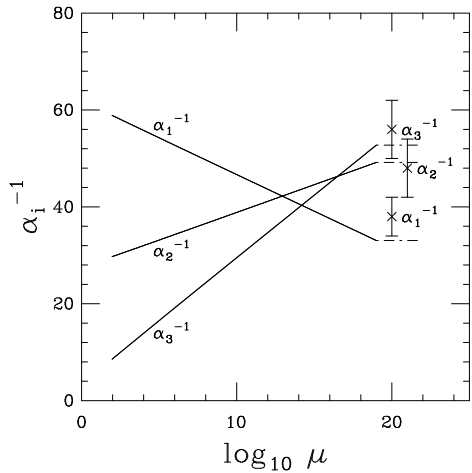,width=6.5cm} }
\caption{Evolution of the Standard Model fine structure constants
$\alpha_i$ ($\alpha_1$ in the SU(5) inspired normalisation) from
the electroweak scale to the Planck scale. The $SMG^3$ gauge group
model predictions for the values at the Planck scale,
$\alpha_i^{-1}(M_{Planck})$, are shown with error bars.}
\label{fig:alphas}
\end{figure}
Actually a major part of this talk will be used to argue that the
application of this MPP may give information about the SM
parameters. In particular we want to use this principle to derive
a value for the top quark Yukawa coupling constant consistent with
experiment and even to solve the hierarchy problem---in the sense
that we get a crude prediction for the electroweak to Planck scale
ratio \cite{fln}. An earlier version of the MPP was first used in
an extension of the SM, in which the gauge bosons occur in three
families analogously to the three quark-lepton families
\cite{bnp,book}. So the gauge group $G$ in this Family Replicated
Gauge Group extension of the SM consists of three copies of the SM
gauge group $SMG$, i.e. $G = SMG^3$ where $SMG = SU(3) \times
SU(2) \times U(1)$. This $SMG^3$ gauge group is supposed to break
down to its diagonal $SMG$ subgroup, identified as the usual SM
gauge group, at some very high energy just below the Planck scale.
According to the MPP, the various $SMG^3$ coupling constants are
supposed to adjust themselves in such a way that there are several
possible vacuum states with the same energy density. The values of
the three SM fine structure constants are thereby predicted at the
Planck scale. The predicted fine struture constants of course do not
unify with a non-simple gauge group but,
as shown in Fig.~\ref{fig:alphas}, they are
consistent with the experimental values extrapolated from the
electroweak scale using the SM renormalisation group equations.
Indeed this calculation was used to predict the number of
quark-lepton families to be three prior to the LEP measurement of
the width of the $Z^0$ gauge boson.

It is most natural to assume that the fundamental gauge, Yukawa
and scalar self-coupling constants should be of order unity.
However the well-known hierarchical structure of the quark-lepton
mass spectrum, with inter-family mass ratios of order 100, is
inconsistent with this assumption, unless one has some extended
theory in which most of the SM Yukawa couplings are effective
rather than fundamental coupling constants. These effective SM
Yukawa coupling constants must be suppressed by some very small
factors. Nevertheless we believe that it is rather difficult for
calculations of effective couplings to give output values which
are not of order unity, if the input numbers are all of order
unity. Therefore if we find one way of generating a very small
number then we suspect Nature may use this way again and again. In
other words if we find some sort of quantity  which is very small
compared to its natural value---we have in mind a Higgs field
vacuum expectation value (VEV)---then it is suggestive to assume
that all the small numbers in the quark-lepton mass spectrum arise
from such quantities. Thus we expect the SM Yukawa coupling
constants to be given by some combinations of interactions, which
must necessarily be proportional to some product of, say, small
Higgs field VEVs.

Of course what we really have in mind is the mass protection
mechanism discussed in Colin Froggatt's lectures, due to the
existence of some (presumably gauge) quantum numbers beyond the SM
which are assigned different values on the left-handed and
right-handed Weyl components of the SM quarks and leptons. By an
appropriate assignment of such chiral quantum numbers to the
quarks and leptons, one can hope to generate effective SM Yukawa
coupling constants suppressed by the appropriate combinations of
Higgs VEVs to be in agreement with experiment. The gauge quantum
numbers of family replicated gauge group models provide attractive
candidates for these mass-protecting chiral quantum numbers.

In section \ref{bound} we present a nice illustrative example of
how one might use the MPP to determine the value of the top quark
SM Yukawa coupling constant. We make the hypothesis that there
exist two approximately degenerate phases of the SM vacuum, in
which the Weinberg Salam Higgs field $\phi_{\sss\rm WS}$ has a VEV 
of the usual electroweak scale order of magnitude. The two phases are
postulated to differ by an effective $6t + 6\overline{t}$ quark
scalar bound state having a zero or non-zero VEV. In section
\ref{scale}, we further postulate the existence of a third vacuum
phase in the pure SM in which $\phi_{\sss\rm WS}$ has a VEV of the order
of the fundamental mass scale, which we take to be the Planck
mass. By requiring this vacuum state to have approximately the
same energy density as the other two vacua, the MPP provides a
solution to the problem of why the electroweak  energy scale is so
tiny compared to the Planck scale; this scale problem is
essentially the well-known hierarchy problem. In section
\ref{masses} we discuss the understanding of the quark-lepton mass
and mixing parameters in the family replicated gauge group model 
\cite{gerry,fnt2}, in which the fundamental couplings are taken to be of order 
unity. Finally we present our conclusions in section \ref{conclusion}.

\section{The Bound State
\label{bound}}

We consider here the possible existence of a $6t + 6\overline{t}$
bound state. Although the dynamics of the binding of such a state
is non-perturbative, it should be relatively insensitive to the
physics above the electroweak scale. So our assumption of pure SM
physics should be rather reliable.

First of all we remark that the virtual exchange of the Weinberg
Salam Higgs particle between two quarks, two anti-quarks or a
quark anti-quark pair yields an attractive force in all cases.
This can be understood by noticing that, even locally, the value
of the Higgs field gives the mass of the quark or anti-quark.
Around a top quark, say, in equilibrium the Higgs field will
reduce the value of its vacuum expectation value a bit, so that
the quark mass and thereby the energy of the quark Higgs
interaction is diminished. In the surroundings of such a quark,
the value of the Higgs field will therefore be pushed down a
little so that the effective or local mass of another quark or
anti-quark will {\em de}crease on entering this region; this means
an attraction between the quarks/anti-quarks. The Higgs field
around a top quark, say, will behave like a Coulomb (or better
Yukawa) field, except for the constant vacuum contribution. Thus
both top ($t$) and anti-top ($\overline{t}$) quarks will attract
each other by Coulomb-like potentials, and we should be able to
roughly calculate the pair binding energy using the Bohr formula
for atomic energy levels.

We now want to argue that, by putting more and more $t$ and
$\overline{t}$ quarks together, the total binding energy could
even compensate for the masses of the quarks and thereby form
a tachyonic bound state. It is then reasonable to expect that a 
new phase could appear due to the formation of a Bose-Einstein 
condensate of these bound state particles. In fact we expect that  
tachyons should really not exist in Nature, but that the vacuum 
condensate should adjust itself to such a density as to bring the 
mass squared of the bound state back to be positive. It is our 
plan here to use the MPP postulate to require the top quark Yukawa 
coupling constant to have precisely that value for which this phase 
transition is reached. If this value turns out to be in agreement 
with the experimental value, it would provide nice phenomenological 
evidence in favour of the MPP hypothesis.

A crude estimate of the binding energy of a collection of $t$ 
and $\overline{t}$ quarks
is given by using the Bohr energy level formula summed over all
{\em pairs} of quarks or anti-quarks. The number of such pair
binding energy contributions increases with the number $n$ of
constituent particles as $n(n-1)/2$, whereas the total rest mass
energy of the constituents ($nm_t$) increases as $n$. So it would
seem that, by combining sufficiently many constituents, the total
binding energy could exceed the rest mass energy of the constituents.
However, we can only put a limited number of $t$ and $\overline{t}$
quarks into the ground state S-wave and, if we use a P-wave, the
pair binding energy is decreased. Nonetheless, one $t$ quark has
two possible spin states and three possible colour states. This
means that $2 \times 3 = 6$ $t$ quarks can be put together in
relative S-waves and, analogously, $2 \times 3 = 6$ $\overline{t}$
quarks in addition. So, in total, we can put $6 + 6 = 12$ such
constituents together in relative S-waves. If we seek to put more
$t$ or $\overline{t}$ quarks together, some of them will have
to go into P-wave or higher energy states and the pair binding energy
will decrease by at least a factor of 4.

\subsection{The Binding Energy Estimate}

In order to make a crude estimate of the binding energy of the
proposed 6 $t$ + 6$\overline{t}$ bound state, we shall first
estimate the binding energy of one of them to the remaining 11
treated as just one collective particle -- ``the nucleus''.
Provided that the radius of the system turns out to be
sufficiently small compared to the Compton wavelength of the
Weinberg Salam Higgs particle, we should be able to use the
well-known Bohr formula for the binding energy of the hydrogen
atom. It is simply necessary to replace the electric charge
$e$ by the top quark Yukawa coupling $\frac{g_t}{\sqrt{2}}$, 
in the normalisation where the kinetic energy term of the Higgs field
$\phi_{\sss\rm WS}$ and the top quark Yukawa interaction in the Lagrangian
density are given by:
\begin{equation}
{\cal{L}} = \frac{1}{2} D_{\mu}\phi_{\sss\rm WS} D^{\mu}\phi_{\sss\rm WS}
+\frac{g_t}{\sqrt{2}}\overline{\Psi}_{tL}\Psi_{tR}\phi_{\sss\rm WS} + h.c.
\end{equation}
In this notation the ``Coulomb'' attraction between two top quarks
is given by the potential:
\begin{equation}
\label{gtpot}
V(r) = -\frac{g_t^2/2}{4\pi r}
\end{equation}
It is easily seen that the attraction between, say, an anti-top quark
and a top quark is given by the same potential (\ref{gtpot}).

We can now estimate the binding energy of one $t$ quark to the system of
the $Z = 11$ remaining particles from the simple hydrogen-like atomic 
energy level formula:
\begin{eqnarray}
E_n & = & -\left(\frac{Zg_t^2/2}{4\pi}\right)^2
\frac{m_t^{reduced}}{2n^2} \\
& = & -\left(\frac{Zg_t^2/2}{4\pi}\right)^2
\frac{Zm_t}{2(Z+1)n^2}
\label{binding}
\end{eqnarray}
where $m_t^{reduced}$ is the redued mass of the top quark and
$n=1$ for the ground state S-wave. In order to obtain the full binding
energy for the 12 particle system, we should add the binding energies
of the other constituents. The simplest crude estimate is just to multiply
the above expression (\ref{binding}) by the number 12 of constituents, taking
into account though that we have thereby double-counted the pairwise
binding contributions. Thus we should really multiply by 6 rather than
by 12. Hence our estimate of the total non-relativistic binding energy
becomes:
\begin{eqnarray}
E_{total \ binding} & = &  6 \left(\frac{11g_t^2/2}{4\pi}\right)^2
\frac{11m_t}{12 \times 2}  \nonumber\\
& = & \left(\frac{11g_t^2/2}{4\pi}\right)^2 \frac{11m_t}{4}
\label{binding1}
\end{eqnarray}

\subsection{u-channel exchange force}

In the above analogy to the hydrogen atom we actually only
considered the t-channel exchange of a Higgs particle between two
t-quarks say -- or between a t-quark and an $\overline{t}$ 
quark -- in the bound state system. We now consider possible Higgs
u-channel exchange contributions. At first one might think there
are none between quarks of different colours, and that is of
course also strictly speaking true. However, if we go to a
formalism in which we fix the colour and spin state of the quarks
so that the $6t+ 6\overline{t}$ state is totally antisymmetric under 
permutations of their colours and spin states, then we can have 
effective u-channel contributions within this formalism. 
Assuming an ansatz state that has the colour and spin degrees of
freedom antisymmetrized in this way, we can formally consider the
quark to be a particle without these degrees of freedom just
including an extra minus sign whenever two quarks are permuted. That 
is to say, under the restriction to just this ansatz wave function
w.r.t. colour and spin, we can effectively consider it as if there
were just one type of quark treated not as a fermion but as a
BOSON. In this formalism there is now clearly place for u-channel
exchanges of even a Higgs particle. In reality it of course
looks strange that a colourless Higgs particle exchange can lead
to the u-channel interaction of say a red and blue quark. But the
point of course is that, before and after the scattering in our
formalism, we project the state onto the subspace with total
antisymmetrization in colour and spin degrees of freedom; so we
let the permutation operators present in the projection operator
help to make the colour exchange, which is needed but not
physically possible for the Higgs particle. So in this formal
sense we then really have u-channel exchange!

Thinking now of the scattering as if it were between identical 
bosonic quarks, the full potential between these quarks is
obtained by adding one term from the t-channel (i.e.~the obvious one)
and one term from the u-channel. For S, D, ... (i.e.~even L)
waves, the u-channel and the t-channel terms between ``bosonic
quarks'' are simply equal in size and add up. This means that, for
instance in the S-wave relevant for our calculation, the
interaction potential between two quarks, due to the collective effect 
of u-channel and t-channel contributions, is just twice as
big as one gets by only calculating the simple Coulomb potential
(meaning only the t-channel contribution).

Even with the above antisymmetrization of the states in spin and
colour, there is no u-channel contribution between the quarks and
the anti-quarks. Really the analogous effect for the quark
anti-quark interaction would rather be s-channel exchange, meaning
the quark anti-quark virtual anihilation diagram. The effect of
such a virtual annihilation could be included but is expected to
contribute less than the u-channel diagrams, because only s-channel 
diagrams having compensating colours on the $t$ and $\overline{t}$ 
quarks can contribute. Furthermore the
s-channel diagram contribution is strongly dependent on the Higgs
mass and the mass of the bound state we are looking for, so we
tend to ignore it at first.

A crude treatment of the u-channel effect is simply to say that
roughly half of the interactions between quarks and anti-quarks in
the bound state get doubled by including the u-channel. We namely
get the doubling for the interactions between anti-quark and
anti-quark, and between quark and quark, but not between
anti-quark and quark. We could therefore crudely say that we
should increase all the Coulomb potentials by an average factor of
$\frac{1}{2} \times 2 + \frac{1}{2} \times 1 = 3/2$. 
This means that we correct for the u-channel by simply replacing the 
previous factor of $g_t^2$ in the Coulomb potential formula 
(\ref{gtpot}) by $\frac{3}{2} \times g_t^2$.
Strictly speaking each quark is only able to interact with 5 other
quarks and not 6, because we do not consider any self-interaction.
So we should really replace $g_t^2$ by $\frac{16}{11}g^2_t$ rather
than by $\frac{3}{2} \times g_t^2$ in the formula (\ref{binding1}) 
for the binding
energy to obtain:
\begin{equation}
 E_{total \ binding} = \left(\frac{2g_t^2}{\pi}\right)^2
 \frac{11m_t}{4}
\label{binding2}
\end{equation}

From consideration of a series of Feynman diagrams or the Bethe-Salpeter
equation for the 12 particle bound state, we would expect that the mass
squared of the bound state, $m_{bound}^2$, should be a more analytic
function of $g_t^2$ than $m_{bound}$ itself. So we now write our crudely
estimated Taylor expansion in $g_t^2$ for the mass {\em squared} of the
bound state:
\begin{eqnarray}
m_{bound}^2 & = & \left(12m_t\right)^2 - 2\left(12 m_t\right)\times
E_{total \ binding} + ...\\
& = & 12m_t^2\left(12 -11 \left(\frac{2g_t^2}{\pi}\right)^2
\frac{2}{4}+ ...\right) \\
& = & \left(12m_t\right)^2\left(1
-\frac{11}{24}\left(\frac{2g_t^2}{\pi}\right)^2
+ ...\right)
\label{expansion}
\end{eqnarray}

\subsection{Estimation of Phase Transition Coupling}

We now assume that, to first approximation, the above formal
Taylor expansion (\ref{expansion}) can be trusted even for
large $g_t$ and with the neglect of higher order terms in the
{\em mass squared} of the bound state. Then the condition that
the bound state should become tachyonic, $m_{bound}^2 < 0$, is
that the top quark Yukawa coupling should be greater than the value
given by the vanishing of equation (\ref{expansion}):
\begin{equation}
0 =  1-\frac{11}{24}\left(\frac{2g_t^2}{\pi}\right)^2 + ...
\label{br0}
\end{equation}
We expect that once the bound state becomes a tachyon, we should
be in a vacuum state in which the effective field, $\phi_{bound}$,
describing the bound state has a non-zero expectation value. Thus
we expect a phase transition just when the bound state mass
squared passes zero\footnote{In fact the phase transition
(degenerate vacuum condition) could easily occur for a small
positive value of $m_{bound}$ and hence a somewhat smaller value
of $g_t^2$.}, which roughly occurs when $g_t$ satisfies the
condition (\ref{br0}) or:
\begin{equation}
g_t|_{phase \ transition} = \sqrt{\frac{\pi}{2}\sqrt{\frac{24}{11}}}
\simeq 1.5
\end{equation}

We have here neglected the attraction due to gluon exchange and the
even smaller electroweak gauge field forces. However the gluon attraction
is rather a small effect compared to the Higgs particle exchange, in
spite of the fact that the QCD coupling $\alpha_s(M_Z) = 0.118$. This
value of the QCD fine structure constant corresponds to a gauge
coupling constant squared value of:
\begin{equation}
g_s^2 = 0.118 \times 4\pi \simeq 1.5
\end{equation}
and an effective gluon-$t\overline{t}$ coupling constant squared of:
\begin{equation}
e_{tt}^2 = \frac{4}{3}g_s^2 \simeq \frac{4}{3} 1.5 \simeq 2.0
\end{equation}
We have to compare this gluon coupling strength $e_{tt}^2 \simeq 2$ with
$Zg_t^2 \simeq 11 \times 1.0$ from the Higgs particle. This leads to a small
gluon exchange correction to the value of $g_t$ at the phase transition:
\begin{equation}
\label{gtop_phase}
 g_t|_{phase\ transition}^{gluon \ corrected} =
 \sqrt{\frac{11\pi}{26}\sqrt{\frac{24}{11}}} \simeq 1.4
\end{equation}
The correction from W-exchange will be smaller than that from gluon
exchange by a multiplicative factor of about
$\frac{\alpha_2(M_Z)}{\alpha_s(M_Z)}
\frac{3}{4} \simeq \frac{1}{5}$,
and the weak hypercharge exchange is further reduced by a factor of
$\sin^2\theta_W$; but overall they will make $g_t|_{phase \ transition}$
a little smaller. Also the s-channel Higgs exchange diagrams will give a 
contribution in the same direction. There are however several effects going in 
the opposite direction, such as the Higgs particle not being truly massless
and that we have over-estimated the concentration of the 11 constituents
forming the ``nucleus''. Furthermore we should consider relativistic 
corrections, but we postpone a discussion of their effects to 
ref.~\cite{fln}. So we should consider our computation to be
rather uncertain at this stage.

We can at least make an estimate of one source of uncertainty, by
considering the effect of using a leading order Taylor expansion in
$g_t^2$ for $m_{bound}$ instead of for $m_{bound}^2$. This would have
led to difference of a factor of 2 in the binding strength and hence
a correction by a factor of the fourth root of 2 in the top quark
Yukawa coupling at the phase boundary; this means a $20\%$ uncertainty
in $g_t|_{phase \ transition}$. Within an uncertainty of this order
of $20\%$, we have a 2 standard deviation difference between the
phase transition (and thus the MPP predicted) coupling, $g_t \simeq 1.4$,
and the measured one, $g_t \simeq 1.0$, corresponding to a physical top
quark mass of about 173 GeV. We thus see that it is quite conceivable,
within our very crude calculations, that the pure SM with the experimental
value of the top quark Yukawa coupling could lie on the boundary
to a new phase; this phase is characterised by a Bose-Einstein condensate
of bound states of the described type, consisting of 6 top quarks and
6 anti-top quarks!

Strictly speaking, if the above MPP scenario is correct, it is not 
obvious in which of the two vacua we live. It may not be so easy 
to detect the postulated bound state condensate even if we were in the phase 
with such a condensate present. There is though a chance that, in 
the presence of the two condensates (of the normal Higgs and the bound 
state fields), the usual Higgs particle could mix with the 
positive mass squared bound state particle. So studies of the 
effects of the Higgs propagator in precision tests of the SM 
could possibly reveal such a mixing. However, in the phase without 
the bound state condensate, we would expect the SM to work 
extremely accurately even if the bound state particle has a low mass. 
This is because it would have very low effective couplings, as the 
amplitude for producing 12 top quark-like particles would be 
very small.  

In the approximation used above the proposed phase transition
caused by a 6t + 6$\bar{t}$ condensate had a phase boundary just
marked by a special value of $g_t$, not depending on the other
coupling constants. This is definitely only an approximation, but
could well be a good approximation. Also one has to ask what would
be meant by the top Yukawa coupling to higher accuracy. As is
well-known one usually defines running Yukawa couplings, which
means their precise definition depends on a renormalization point
(= scale parameter) $\mu$. This means that the value we have
derived above, see eq.~(\ref{gtop_phase}), for $g_t$ at the phase
transition has to be assigned a scale $\mu$ at which it should be
valid. Since the energy scales were all crudely of the order of
the electroweak scale $\mu_{weak} \sim 100$ GeV, it is rather 
obvious that we found 
$g_t|_{phase\ transition}^{gluon \ corrected}( \mu_{weak}) \simeq 1.4$.

This possibility of a new phase at the electroweak scale provides a very
clean test of our MPP hypothesis. Since it only involves SM physics at the
electroweak scale, it should be possible to make more accurate and
reliable calculations of the value of the top quark Yukawa coupling
constant at the phase transition. This would then test the MPP
hypothesis independent of any further speculative hypotheses beyond the
SM.

\section{Solution of the Scale Problem by MPP}
\label{scale}

We shall now see how the above suggested bound state can be used together 
with our principle of degenerate vacua to solve what we could call the 
``scale problem''. The scale problem refers to the question of why the 
electroweak scale $\mu_{weak}$ is so extremely small compared to the 
fundamental scale, say the Planck scale $\Lambda_{Planck} \sim 10^{19}$ GeV. 
This problem is of course connected to, but really not identical to, the 
``hierarchy problem'' of how to avoid having to shuffle around the quadratic 
divergences in the SM Higgs mass squared $M_H^2$, which arise as one goes 
from one order in perturbation theory to the next. In fact we reformulate 
the fine-tuning involved in solving the scale problem into the fine-tuning 
of several vacuum energy densities to be approximately equal and close 
to zero. We thereby unify several successful fine-tuning predictions into 
a single fine-tuning principle -- the Multiple Point Principle. Actually
we shall use the MPP to tune the value of the running top quark Yukawa 
coupling {\it both} at the fundamental scale and at the electroweak scale, 
but to somewhat different numerical values. Since running couplings vary 
logarithmically with scale, such a requirement can easily give an 
exponentially large scale ratio and thereby solve the scale problem.

\subsection{Two degenerate minima in the effective Higgs potential}

Now let us briefly review a different example which, prior to the
bound state story presented above, was the MPP application least
dependent on guessing the physics beyond the SM. We investigated
\cite{fn2} the MPP-requirement that the SM effective potential for
the Weinberg Salam Higgs field should have two degenerate minima.
This requirement leads to the condition that our vacuum is barely
stable, so that the top quark and SM Higgs masses should lie on
the SM vacuum stability curve \cite{sher,colin}. Using the measured
value of the top quark mass, this MPP condition predicts a Higgs
mass of about 135 GeV. As a suggestive hypothesis, we added the
assumption that the VEV in the second minimum should be of the order
of the ``fundamental'' energy scale $\phi_{vac\; 2} \sim \mu_{fundamental}$,
which we took to be the Planck scale $\Lambda_{Planck}$.
This gave, with surprisingly good accuracy, our predicted combination of
(pole) masses:
\begin{equation}
\label{tophiggs}
M_{t} = 173 \pm 5\ \mbox{GeV} \quad M_{H} = 135 \pm 9\ \mbox{GeV}
\end{equation}
We could say that, with the latter assumption regarding the VEV in
the second minimum, the numerical coincidence between the
predicted and measured top quark masses provided support for MPP
even at that time (1995). Later, with Yasutaka Takanishi, we
discussed the possibility \cite{fnt} that the MPP condition should
predict mass values corresponding to metastability of the vacuum
rather than true vacuum stability. This would give instead a Higgs
mass prediction of $122 \pm 11$ GeV, close to the LEP lower bound
of 115 GeV \cite{pdg}. We should presumably not really take the
MPP predictions to be more accurate than to the order of magnitude
of the variation between the metastability and stability bounds.
However we definitely predict a light Higgs mass in this range, as
seems to be in agreement with indirect estimates of the SM Higgs
mass from precision data \cite{pdg}.

\subsection{The three vacuum degeneracy assumption}

We seem to have here two successful examples of MPP-predictions: The
one from 1995 concerning the two minima in the Higgs effective
potential and the bound state case discussed in section
\ref{bound}. It is obvious that if we want to make use of both of
them, then we should in reality postulate the existence of at
least three vacua with the same energy density. There should
namely be at least two of them deviating by having the $6t +
6\overline{t}$ bound state condensate or not. Furthermore there
should be two vacua which deviate essentially just by the Weinberg
Salam Higgs field having the small ($246$ GeV) or the large ($\sim
\mu_{fundamental}$) vacuum expectation value. Such a several
degenerate vacua picture is of course exactly what MPP is supposed
to mean.

\subsection{The running top Yukawa coupling from MPP for two different scales}

Now we turn to the scale problem of why the electroweak scale
$\mu_{weak}$ is so tiny compared to the fundamental scale
$\mu_{fundamental}$, which is essentially the hierarchy problem.
For the purpose of attacking this scale problem, we shall argue
that we may look at the MPP requirement that there be two
degenerate minima in the Higgs effective potential as telling us
about the value of the running top Yukawa coupling at the
fundamental scale.

As discussed in Colin's lectures \cite{colin}, rather simple
conditions are obtained to leading order (which is an excellent
approximation) for the running couplings at this fundamental
scale, where we imagine the higher one of the minima in the Higgs
effective potential to be $\phi_{vac\; 2} = \mu_{fundamental}$,
namely:
\begin{eqnarray}
\lambda(\phi_{vac\; 2})& = & 0 \\
\beta_{\lambda}(\lambda(\phi_{vac\; 2}), g_t(\phi_{vac\; 2}),
\alpha_1, \alpha_2) & = & 0.
\end{eqnarray}
Here $\lambda(\mu)$ is the Higgs self-coupling constant and 
$\beta_{\lambda}=\frac{d\lambda}{d\ln\mu}$ is the corresponding beta 
function. We assume that $\phi_{vac\; 2} \gg \phi_{vac\; 1} = 246$ GeV 
(although not necessarily by many orders of magnitude) so that, at the
second minimum, the mass term $m^2 |\phi_{\sss\rm WS}|^2$ is negligible
and the renormalisation
group improved efective potential $V_{eff}(\phi_{\sss\rm WS})$ is
dominated by the term $\lambda(\phi_{\sss\rm WS})|\phi_{\sss\rm WS}|^4$. 
This leads to the expression
\begin{equation}
\label{top2}
 g_t = \left(\pi^2\left[\alpha_1^2 + 2\alpha_1\alpha_2
 + 3\alpha_2^2\right]\right)^{1/4}
\end{equation}
for the running top quark Yukawa coupling constant in terms of the
electroweak fine structure constants $\alpha_1(\mu)$ and
$\alpha_2(\mu)$ at the scale $\mu = \phi_{vac\; 2}$. We need here to
input the experimental values of the fine structure constants.
However, we note that the value of the right-hand side of
eq.~(\ref{top2}) is rather insensitive to the scale $\mu$, varying
by approximately $10\%$ between $\mu = 246$ GeV and $\mu
=\Lambda_{Planck}$. So we obtain the value
\begin{equation}
\label{gtop_fundamental}
g_t(\mu_{fundamental}) \simeq 0.4.
\end{equation}

From our three SM vacuum degeneracy assumption, we have obtained
MPP predictions for the top quark Yukawa coupling at the electroweak
scale, eq.~(\ref{gtop_phase}), and the fundamental scale,
eq.~(\ref{gtop_fundamental}). So we can calculate an MPP prediction
for the ratio of these scales $\mu_{weak}/\mu_{fundamental}$, using
the renormalisation group equations; the fine structure constants
contributing to the running are considered as given. Although the
two MPP predicted values of the top quark Yukawa coupling $g_t(\mu)$
are of order unity, differing by a factor of order $e$, it should
be noticed that the beta function for $g_t$
\begin{equation}
 \beta_{g_t} = \frac{dg_t}{d\ln\mu} =
 \frac{g_t}{4\pi}\left(\frac{9}{2}
 \cdot\frac{g_t^2}{4\pi} - 8\alpha_3 - \frac{9}{4}\alpha_2
 - \frac{17}{12}\alpha_1\right)
\label{betatop}
\end{equation}
is relatively small. So the logarithm of the scale ratio
$\ln \mu_{fundamental}/\mu_{weak}$ needed to generate the required
amount of renormalisation group running of $g_t({\mu})$ must be a
large number. Hence the scale ratio itself must be huge, essentially
providing a solution to the hierarchy problem! So we claim in this way
to explain why the weak scale $\mu_{weak}$ is so low compared to, say,
the Planck scale $\Lambda_{Planck}$, although in practice MPP only
predicts the order of magnitude of the logarithm of the scale ratio.
However, if we assume that a more accurate MPP calculation of
$g_t|_{phase \ transition}(\mu_{weak})$ will turn out to agree with the
experimental value $g_t(\mu_{weak}) \simeq 1.0$, we can use our 1995
result \cite{fn2,colin} to give the MPP prediction
\begin{equation}
\frac{\mu_{weak}}{\mu_{fundamental}} \sim 10^{-17}
\end{equation}
which is consistent with identifying the fundamental scale with the
Planck scale.

\section{Quark-lepton masses and mixings}
\label{masses}

Having discussed the MPP favoured values for the top quark and
SM Higgs masses, eq.~(\ref{tophiggs}), we now turn to how the other
quark-lepton masses are understood in the Family Replicated Gauge
Group Model, which was briefly described in section \ref{introduction}
in connection with an MPP prediction of the fine structure constants.
In this model, the gauge group is replicated, one copy for each
quark-lepton proto-family. So in contrast to grand unified theories,
the model naturally contains many chiral gauge charges, which
distinguish the families and can be used for mass protection \cite{colin}.
The fundamental couplings in the model are assumed to be of order
unity and the Higgs field VEVs, which break the replicated gauge group
down to the SM gauge group (SMG), provide the small parameters necessary to
generate the ``small hierarchy'' of fermion masses (e.g.~to explain why 
the $\tau$-lepton mass is much bigger than the electron mass).

As illustrated in Fig.~1, the MPP prediction for the values of the
SM fine structure constants agrees well with experiment in the
family replicated $SMG^3$ gauge group model, when broken down to
the diagonal $SMG$ close to the Planck scale $\Lambda_{Planck}$, which is
taken to be the fundamental scale of the theory. In fact we typically
take the corresponding Higgs field VEVs responsible for the breaking
\begin{eqnarray}
& & SMG^3 \rightarrow SMG_{diag} \\
 SMG_{diag} & = & \left\{ (g,g,g)|g \in SMG \right\} 
 \subseteq SMG \otimes SMG \otimes SMG
\end{eqnarray}
to be of order $\Lambda_{Planck}/10$. For this type of breaking it is
rather easy, in the weak field approximation, to obtain the relation
between the effective fine structure constant $\alpha_i^{diag}$ for
the diagonal non-abelian subgroup in terms of the corresponding fine 
structure constants $\alpha_i^j$, where $j=1,2,3$ labels the families, 
in the family replicated gauge groups:
\begin{equation}
 \frac{1}{\alpha_i^{diag}} = \sum_{j=1}^3 \frac{1}{\alpha_i^j} 
 \qquad i=2,\ 3.
\label{diagrel}
\end{equation}
The situation is more complicated for the abelian 
groups, because it is possible to have gauge invariant 
cross-terms between the different $U(1)$ groups in 
the Lagrangian density, such as:
\begin{equation}
\frac{1}{4g_{1,2}^2} F_{\mu\nu}^{1}(x) F_{2}^{\mu \nu}(x)
\end{equation}
between the first and second proto-families. So the full 
expression for the $U(1)$ case becomes:
\begin{equation}
 \frac{1}{\alpha_1^{diag}} = \sum_{j=1}^3 \frac{1}{\alpha_1^j} 
 + \sum_{(j<k)=1}^3 \frac{1}{\alpha_{j,k}}.
\label{diagabel}
\end{equation}
where $\alpha_{j,k}=g_{j,k}^2/4\pi$.  

According to the MPP, the 
$SMG^3$ coupling constants should be fixed so as to ensure the 
existence of many vacuum states with the same energy density; 
in the Euclideanised version of the theory, there is a 
corresponding phase transition. So if several vacua 
are degenerate, there is a multiple point. The couplings 
at the multiple point have been calculated in 
lattice gauge theory \cite{lattice} for the groups $SU(3)$, 
$SU(2)$ and $U(1)$ separately. Here we imagine that the 
lattice has a truly physical significance in providing 
a cut-off for our model at the Planck scale. An alternative 
viewpoint, to taking the lattice seriously as really existing, 
is to assume the existence of very heavy monopoles \cite{larisa}. 
The SM fine structure constants correspond to those of 
the diagonal subgroup of the $SMG^3$ group and,
for the non-abelian groups, this gives:
\begin{equation}
\alpha_i(\Lambda_{Planck}) = \frac{\alpha_i^{MPP}}{3} 
\qquad i=2,\ 3 
\end{equation}
since $\alpha_i^j = \alpha_i^{MPP}$ for all three families. 
These are the MPP predictions shown in Fig.~\ref{fig:alphas}. 
For the abelian case we took the suggestive estimate 
$\alpha_{j,k} = \alpha_1^{MPP}$ and hence 
\begin{equation}
\alpha_1(\Lambda_{Planck}) = \frac{\alpha_1^{MPP}}{6} 
\end{equation}  
although this is less reliable than the non-abelian case.

\subsection{The ``maximal'' family replicated gauge group}

The complicated hierarchy of quark-lepton masses suggests the
existence of chiral (gauge) quantum numbers, which are rather
different for the various SM Weyl fields. This may well be most
easily realised by taking a large gauge group. Now the only part
of the gauge group beyond the SM to which we can hope to have
phenomenological access is that part which transforms the
physically observable Weyl fields. This means, first of all, the 
known 45 SM Weyl particles. If we optimistically include -- as in 
$SO(10)$ inspired models -- three families of right-handed (see-saw) 
neutrinos, then we end up with 48 ``phenomenologically accessible'' 
Weyl fields. The relevant gauge group should be a subgroup of 
all the unitary transformations of the phenomenologically 
accessible Weyl fields. It should therefore be embedded in 
$U(45)$ if we include just the SM fermions or in $U(48)$ if the 
right-handed neutrinos are also included.

It has been argued that global (i.e.~non-gauge) symmetries 
are broken by quantum gravity effects. So we shall assume that 
all the mass-protecting symmetries are gauge symmetries and thus
require that our gauge group should have no gauge and no mixed 
anomalies. These requirements mean that neither the gauge symmetry 
nor the symmetry needed for the inclusion of gravity will be broken 
by radiative loop corrections in the theory. It can, in fact, 
easily happen that the gauge symmetry is broken by so-called 
triangle diagrams, consisting of a Weyl fermion loop with three external 
gauge bosons. In realistic theories, such as the SM, it is only by 
a carefully organised cancellation between different Weyl fields 
that the collective effect of gauge charge violation is brought to 
zero. An idea about the physics underlying anomalies is provided 
by the remark that the gauge fields can drive some Weyl particles 
carrying a gauge charge up from the Dirac sea into positive energy 
states. These Weyl particles look as if they were produced just 
from the gauge field and the gauge charge conservation appears to have been  
broken; one then talks about an anomaly. Indeed if one attempted to 
gauge the full $U(45)$ extension of the SM group, then triangle diagrams 
would lead to gauge symmetry breaking and the $U(45)$ theory would be 
unacceptable, at least not without supplementing it with a lot more 
Weyl fermions. Similarly, with three right-handed neutrinos, the 
full $U(48)$ gauge group extension is not allowed.

This discussion raises the question of which is the largest subgroup 
of $U(45)$ extending the SM that is anomaly free. The answer is  
$SU(5)^3$ \cite{rajpoot} supplemented with an abelian flavour 
group, i.e. $SU(5)^3\otimes U(1)_f$. However if one wants to avoid 
the simple $SU(5)$ mass predictions, such as $m_d/m_s = m_e/m_{\mu}$,   
without complicating the Higgs representations, one should take a 
non-unifying group. So we shall require that none of the SM irreducible 
representations of Weyl fields become united into larger irreducible 
representations of the extended gauge group. With this latter 
requirement the maximal allowed subgroup of $U(45)$ becomes 
\cite{trento} $SMG^3 \otimes U(1)_f$.

Similarly, including three right-handed neutrinos, the biggest anomaly 
free subgroup of $U(48)$ is $SO(10)^3$, i.e.~the family replicated 
$SO(10)$ model \cite{ling}. Again we are interested in avoiding 
unification and leaving the SM irreducible representations of Weyl 
fields as irreducible under the extended group. So we reduce the 
$SO(10)^3$ group to the group 
\begin{equation}
\label{frgg}
G = (SMG \otimes U(1)_{B-L})^3 = SU(3)^3\otimes SU(2)^3 \otimes U(1)^6.
\end{equation}
This family replicated group G is the maximal anomaly free subgroup 
of $U(48)$, extending the SM plus three right-handed neutrinos, without 
unifying particles not already in the same SM irreducible representations. 
We shall now consider a five parameter fit to the quark-lepton masses and 
mixing angles in this model \cite{fnt2}.  

\subsection{The $(SMG \times U(1)_{B-L})^3$ Model}

In this family replicated gauge group model we introduce the 
extended gauge group $G = (SMG \times U(1)_{B-L})^3$,
where the three copies of the SM gauge group are supplemented by
an abelian $(B-L)$ (= baryon number minus lepton number) gauge
group for each family. There are 6 abelian
gauge charges -- the weak hypercharge $y/2$ and the $(B-L)$ charge for 
the 3 different families -- which are responsible for generating 
the fermion mass hierarchy and we list their values in Table
\ref{Table1} for the 48 Weyl proto-fermions in the model.
\begin{table}[!ht]
\caption{All $U(1)$ quantum charges -- the three weak hypercharges 
$y_i/2$ and the three $(B-L)$-charges $(B-L)_i$ -- for the proto-fermions 
in the $(SMG \times U(1)_{B-L})^3$ model.}  
\label{Table1}
\begin{center}
\begin{tabular}{|c||c|c|c|c|c|c|} \hline
& $y_1/2$& $y_2/2$ & $y_3/2$ & $(B-L)_1$ & $(B-L)_2$
& $(B-L)_3$ \\ \hline\hline
$u_L,d_L$ &  $\frac{1}{6}$ & $0$ & $0$ & $\frac{1}{3}$ & $0$ & $0$ \\
$u_R$ &  $\frac{2}{3}$ & $0$ & $0$ & $\frac{1}{3}$ & $0$ & $0$ \\
$d_R$ & $-\frac{1}{3}$ & $0$ & $0$ & $\frac{1}{3}$ & $0$ & $0$ \\
$e_L, \nu_{e_{\sss L}}$ & $-\frac{1}{2}$ & $0$ & $0$ & $-1$ & $0$ & $0$ \\
$e_R$ & $-1$ & $0$ & $0$ & $-1$ & $0$ & $0$ \\
$\nu_{e_{\sss R}}$ &  $0$ & $0$ & $0$ & $-1$ & $0$ & $0$ \\ \hline
$c_L,s_L$ & $0$ & $\frac{1}{6}$ & $0$ & $0$ & $\frac{1}{3}$ & $0$ \\
$c_R$ &  $0$ & $\frac{2}{3}$ & $0$ & $0$ & $\frac{1}{3}$ & $0$ \\
$s_R$ & $0$ & $-\frac{1}{3}$ & $0$ & $0$ & $\frac{1}{3}$ & $0$\\
$\mu_L, \nu_{\mu_{\sss L}}$ & $0$ & $-\frac{1}{2}$ & $0$ & $0$ &
$-1$ &
$0$\\ $\mu_R$ & $0$ & $-1$ & $0$ & $0$  & $-1$ & $0$ \\
$\nu_{\mu_{\sss R}}$ &  $0$ & $0$ & $0$ & $0$ & $-1$ & $0$ \\
\hline
$t_L,b_L$ & $0$ & $0$ & $\frac{1}{6}$ & $0$ & $0$ & $\frac{1}{3}$ \\
$t_R$ &  $0$ & $0$ & $\frac{2}{3}$ & $0$ & $0$ & $\frac{1}{3}$ \\
$b_R$ & $0$ & $0$ & $-\frac{1}{3}$ & $0$ & $0$ & $\frac{1}{3}$\\
$\tau_L, \nu_{\tau_{\sss L}}$ & $0$ & $0$ & $-\frac{1}{2}$ & $0$ &
$0$ &
$-1$\\ $\tau_R$ & $0$ & $0$ & $-1$ & $0$ & $0$ & $-1$\\
$\nu_{\tau_{\sss R}}$ &  $0$ & $0$ & $0$ & $0$ & $0$ & $-1$ \\
\hline 
\end{tabular}
\end{center}
\end{table}

We must now consider the Higgs fields responsible for the breakdown 
of the family replicated gauge group (\ref{frgg}) to the SM group.
We have been able to construct a set of Higgs fields and to assign
VEVs to them, so as to fit all the quark-lepton masses and mixing 
angles, under the assumption that all the fundamental couplings are of 
order unity and the fundamental masses are of order the Planck mass 
$\Lambda_{Planck}$. In this model, we assume that there is a rich 
spectrum of Planck scale particles. Such Planck mass particles of 
almost any wanted quantum numbers are important in the model, in as far 
as they provide intermediate states in the transitions from left-handed 
to right-handed quarks and leptons needed to produce their masses 
\cite{colin,fn1}. The transitions between the intermediate fermion 
states are caused by Yukawa couplings to Higgs fields with non-zero 
VEVs. In this way we express the orders of magnitude of the quark-lepton 
mass matrix elements (in units of the Weinberg Salam Higgs field VEV) 
as products of Higgs field VEVs measured in Planck units. So it is the 
smallness of the Higgs field VEVs compared to the fundamental scale, as 
manifested in the massive particle propagators, that is responsible for 
suppressing quark-lepton masses.

The abelian gauge quantum
numbers of the system of Higgs fields for the $(SMG \times
U(1)_{B-L})^3$ model are given in Table \ref{qc}.
%%%%%%%%%%%%%%%%%%%%%%%%%%%%%%%%%%%%%%%%%%%%%%%%%%%%%%%
\begin{table}[!th]
\caption{All $U(1)$ quantum charges of the Higgs fields in the
$(SMG \times U(1)_{B-L})^3$ model.} \vspace{3mm} \label{qc}
\begin{center}
\begin{tabular}{|c||c|c|c|c|c|c|} \hline
& $y_1/2$& $y_2/2$ & $y_3/2$ & $(B-L)_1$ & $(B-L)_2$
& $(B-L)_3$ \\ \hline\hline
$\omega$ & $\frac{1}{6}$ & $-\frac{1}{6}$ & $0$ & $0$ & $0$ & $0$\\
$\rho$ & $0$ & $0$ & $0$ & $-\frac{1}{3}$ & $\frac{1}{3}$ & $0$\\
$W$ & $0$ & $-\frac{1}{2}$ & $\frac{1}{2}$ & $0$ & $-\frac{1}{3}$
& $\frac{1}{3}$ \\
$T$ & $0$ & $-\frac{1}{6}$ & $\frac{1}{6}$ & $0$ & $0$ & $0$\\
$\phi_{\sss WS}$ & $0$ & $\frac{2}{3}$ & $-\frac{1}{6}$ & $0$
& $\frac{1}{3}$ & $-\frac{1}{3}$ \\
$\phi_{\sss SS}$ & $0$ & $1$ & $-1$ & $0$ & $2$ & $0$ \\
\hline
\end{tabular}
\end{center}
\end{table}
%%%%%%%%%%%%%%%%%%%%%%%%%%%%%%%%%%%%%%%%%%%%%%%%%%%%%%%
We assume that, like the quark and lepton fields, the Higgs 
fields belong to singlet or fundamental representations of 
all the non-abelian groups. Then, by imposing the usual SM 
charge quantisation rule for each of the $SMG$ factors, the 
non-abelian representations are determined from the weak 
hypercharge quantum numbers $y_i/2$.

With the system of quantum numbers
in Table~\ref{qc} one can easily evaluate, for a given mass matrix
element, the numbers of Higgs field VEVs of the different types
needed to perform the transition between the corresponding left-
and right-handed Weyl fields. The results of calculating the
products of Higgs fields needed, and thereby the orders of
magnitude of the mass matrix elements in our model, are presented
in the following mass matrices (where, for clarity, we distinguish
between Higgs fields and their hermitian conjugates):

\noindent the up-type quarks:
\begin{eqnarray}
M_{\sss U} \simeq \frac{\sVEV{(\phi_{\sss\rm WS})^\dagger}}
{\sqrt{2}}\hspace{-0.1cm} \left(\!\begin{array}{ccc}
        (\omega^\dagger)^3 W^\dagger T^2
        & \omega \rho^\dagger W^\dagger T^2
        & \omega \rho^\dagger (W^\dagger)^2 T\\
        (\omega^\dagger)^4 \rho W^\dagger T^2
        &  W^\dagger T^2
        & (W^\dagger)^2 T\\
        (\omega^\dagger)^4 \rho
        & 1
        & W^\dagger T^\dagger
\end{array} \!\right)\label{M_U}
\end{eqnarray}
\noindent the down-type quarks:
\begin{eqnarray}
M_{\sss D} \simeq \frac{\sVEV{\phi_{\sss\rm WS}}}{\sqrt{2}}
\hspace{-0.1cm} \left (\!\begin{array}{ccc}
        \omega^3 W (T^\dagger)^2
      & \omega \rho^\dagger W (T^\dagger)^2
      & \omega \rho^\dagger T^3 \\
        \omega^2 \rho W (T^\dagger)^2
      & W (T^\dagger)^2
      & T^3 \\
        \omega^2 \rho W^2 (T^\dagger)^4
      & W^2 (T^\dagger)^4
      & W T
                        \end{array} \!\right) \label{M_D}
\end{eqnarray}
\noindent %
the charged leptons:
\begin{eqnarray}
M_{\sss E} \simeq \frac{\sVEV{\phi_{\sss\rm WS}}}{\sqrt{2}}
\hspace{-0.1cm} \left(\hspace{-0.1 cm}\begin{array}{ccc}
    \omega^3 W (T^\dagger)^2
  & (\omega^\dagger)^3 \rho^3 W (T^\dagger)^2
  & (\omega^\dagger)^3 \rho^3 W^4 (T^\dagger)^5  \\
    \omega^6 (\rho^\dagger)^3  W (T^\dagger)^2
  &   W (T^\dagger)^2
  &   W^4 (T^\dagger)^5 \\
    \omega^6 (\rho^\dagger)^3  (W^\dagger)^2 T^4
  & (W^\dagger)^2 T^4
  & WT
\end{array} \hspace{-0.1cm}\right) \label{M_E}
\end{eqnarray}
\noindent the Dirac neutrinos:
\begin{eqnarray}
M^D_\nu \simeq \frac{\sVEV{(\phi_{\sss\rm WS})^\dagger}}
{\sqrt{2}}\hspace{-0.1cm} \left(\hspace{-0.1cm}\begin{array}{ccc}
        (\omega^\dagger)^3 W^\dagger T^2
        & (\omega^\dagger)^3 \rho^3 W^\dagger T^2
        & (\omega^\dagger)^3 \rho^3 W^2 (T^\dagger)^7 \\
        (\rho^\dagger)^3 W^\dagger T^2
        &  W^\dagger T^2
        & W^2 (T^\dagger)^7 \\
        (\rho^\dagger)^3 (W^\dagger)^4 T^8
        &   (W^\dagger)^4 T^8
        & W^\dagger T^\dagger
\end{array} \hspace{-0.1 cm}\right)\label{Mdirac}
\end{eqnarray}
\noindent %
and the Majorana (right-handed) neutrinos:
\begin{eqnarray}
M_R \simeq \sVEV{\phi_{\sss\rm SS}}\hspace{-0.1cm} \left
(\hspace{-0.1 cm}\begin{array}{ccc} (\rho^\dagger)^6 T^6 &
(\rho^\dagger)^3 T^6
& (\rho^\dagger)^3 W^3 (T^\dagger)^3 \\
(\rho^\dagger)^3 T^6
& T^6 & W^3 (T^\dagger)^3 \\
(\rho^\dagger)^3 W^3 (T^\dagger)^3 & W^3 (T^\dagger)^3 & W^6
(T^\dagger)^{12}
\end{array} \hspace{-0.1 cm}\right ) \label{Mmajo}
\end{eqnarray}

As can be seen from Table~\ref{qc}, the fields $W$ and $T$ only have 
non-trivial quantum numbers with respect to the second and third 
families and, together with $\phi_{\sss\rm WS}$ (or $\phi_{\sss SS}$), 
they determine the orders 
of magnitude of the 2-3 submatrix elements in all the mass matrices 
(\ref{M_U}) - (\ref{Mmajo}). Also the fields $\omega$ and $\rho$ only 
have non-trivial quantum numbers with respect to the first
and second families. This choice of quantum numbers makes it
possible to express a fermion mass matrix element involving the
first family in terms of the corresponding element involving the
second family, by the inclusion of an appropriate product of
powers of $\rho$ and $\omega$. We remark that corresponding diagonal 
elements are the same (apart from possible charge conjugation) for 
all the Dirac mass matrices (\ref{M_U}) - (\ref{Mdirac}). 
Consequently the top quark mass is 
given by an off-diagonal element $\left(M_U\right)_{32}$. We note 
that the Higgs field $\phi_{\sss SS}$ is introduced to break the 
diagonal $(B-L)$ gauge symmetry and sets the see-saw scale for the 
right-handed neutrinos. The quantum numbers of $\phi_{\sss SS}$ are 
rather large, which may indicate that it should be replaced by the 
repetitive use of a field having quantum numbers equal to a fraction 
(say 1/3 or 1/6) of the $\phi_{\sss SS}$ quantum numbers in Table~\ref{qc}.
 
The light neutrino mass matrix -- effective left-left
transition Majorana mass matrix -- can be obtained via the see-saw
mechanism~\cite{seesaw}:
\begin{equation}
  \label{eq:meff}
  M_{\rm eff} \! \approx \! M^D_\nu\,M_R^{-1}\,(M^D_\nu)^T\nn.
\end{equation}
with an appropriate renormalisation group running from the Planck
scale to the see-saw scale and then to the electroweak scale. The
experimental quark and lepton masses and mixing angles in Table
\ref{convbestfit} can now be fitted, by varying just 5 Higgs field
VEVs and averaging over a set of complex order unity random
numbers, which multiply all the independent mass matrix elements.
The best fit is obtained with the following values for the VEVs:
\begin{eqnarray}
\label{eq:VEVS} &&\sVEV{\phi_{\sss SS}}=5.25\times10^{15}~\GeV\nn,
\nn\sVEV{\omega}=0.244\nn, \nn\sVEV{\rho}=0.265\nn,\nonumber\\
&&\nn\sVEV{W}=0.157\nn, \nn\sVEV{T}=0.0766\nn,
\end{eqnarray}
where, except for the Higgs field $\sVEV{\phi_{\sss SS}}$, the
VEVs are expressed in Planck units. The resulting 5 parameter
order of magnitude fit, with an LMA-MSW solution to the solar
neutrino problem, is presented in Table \ref{convbestfit}.  In 
addition the natural expectation in our model, that the CP violating 
phase in the quark (CKM) mixing matrix $V$ is of order unity, is 
in agreement with phenomenological fits to the unitarity triangle 
\cite{vysotsky}

%%%%%%%%%%%%%%%%%%%%%%%%%%%%%%%%%%%%%%%%%%%%%
\begin{table}[!t]
\caption{Best 5 parameter fit to conventional experimental data. 
All masses are running masses at $1~\GeV$ except the top quark mass 
which is the pole mass.}
\begin{displaymath}
\begin{array}{|c|c|c|}
\hline
 & {\rm Fitted} & {\rm Experimental} \\ \hline
m_u & 4.4~\MeV & 4~\MeV \\
m_d & 4.3~\MeV & 9~\MeV \\
m_e & 1.6~\MeV & 0.5~\MeV \\
m_c & 0.64~\GeV & 1.4~\GeV \\
m_s & 295~\MeV & 200~\MeV \\
m_{\mu} & 111~\MeV & 105~\MeV \\
M_t & 202~\GeV & 180~\GeV \\
m_b & 5.7~\GeV & 6.3~\GeV \\
m_{\tau} & 1.46~\GeV & 1.78~\GeV \\
V_{us} & 0.11 & 0.22 \\
V_{cb} & 0.026 & 0.041 \\
V_{ub} & 0.0027 & 0.0035 \\ \hline
\Delta m^2_{\odot} & 9.0 \times 10^{-5}~\eV^2 &  5.0 \times 10^{-5}~\eV^2 \\
\Delta m^2_{\rm atm} & 1.7 \times 10^{-3}~\eV^2 &  2.5 \times 10^{-3}~\eV^2\\
\tan^2\theta_{\odot} &0.26 & 0.34\\
\tan^2\theta_{\rm atm}& 0.65 & 1.0\\
\tan^2\theta_{\rm chooz}  & 2.9 \times 10^{-2} & \sleq~2.6 \times 10^{-2}\\
%\hline\hline
%\mbox{\rm g.o.f.} &  3.63 & - \\
\hline
\end{array}
\end{displaymath}
\label{convbestfit}
\end{table}
%%%%%%%%%%%%%%%%%%%%%%%%%

Transforming from $\tan^2\theta$ variables to $\sin^22\theta$
variables, our predictions for the neutrino mixing angles become:
\begin{equation}
  \label{eq:sintan}
 \sin^22\theta_{\odot} = 0.66\nn, \quad
 \sin^22\theta_{\rm atm} = 0.96\nn, \quad
 \sin^22\theta_{\rm chooz} = 0.11\nn.
\end{equation}
Note that our fit to the CHOOZ mixing angle lies close to the
$2\sigma$ Confidence Level experimental bound. We also give here
our predicted hierarchical left-handed neutrino masses ($m_i$) and
the right-handed neutrino masses ($M_i$) with mass eigenstate
indices ($i=1,2,3$):
\begin{equation}
m_1 =  1.4\times10^{-3}~~\eV\nn, \quad M_1 =
1.0\times10^{6}~~\GeV\nn, \label{eq:neutrinomass1}
\end{equation}
\begin{equation}
m_2 =  9.6\times10^{-3}~~\eV\nn, \quad M_2 =
6.1\times10^{9}~~\GeV\nn, \label{eq:neutrinomass2}
\end{equation}
\begin{equation}
m_3 =  4.2\times10^{-2}~~\eV\nn, \quad M_3 =
7.8\times10^{9}~~\GeV\nn. \label{eq:neutrinomass3}
\end{equation}

We see from Table \ref{convbestfit} that the data are fit reasonably well 
order of magnitudewise, actually even with the expected accuracy 
\cite{douglas}. It should also be remarked that the heavy right-handed 
neutrinos are good candidates for generating the baryon asymmetry in 
the Universe, via their CP violating decays and leptogenesis 
\cite{fukugita} in the era when the temperature of the Universe was of 
the same order of magnitude as their masses.

\section{Conclusion}
\label{conclusion}

As a first step towards predicting the values of the SM parameters and, in 
particular, the ratio of the electroweak scale to the Planck scale, we have 
put forward the Multiple Point Principle (MPP). This MPP states that there 
exist many different vacuum states having about the same energy density 
(more precisely all having approximately zero energy density or cosmological 
constant). We also proposed the hypothesis that all the ``fundamental'' 
couplings are of order unity.

One would normally expect that the determination of coupling constants from 
some principle, such as the MPP, would tend to give values of order 
unity. However this order of unity result is not completely general, as was 
illustrated by the MPP mechanism we put forward to explain the very large 
scale ratio between the fundamental and electroweak scales. In this example 
the big scale ratio came out successfully, at least within the uncertainties 
of the non-perturbative calculations, as the exponential of a quantity 
involving essentially the inverse of the SM fine structure constants. This 
application of the MPP to the pure SM determined the approximate value of the 
running top quark Yukawa coupling $g_t(\mu)$ at two different scales -- namely 
at the electroweak scale $\mu_{weak}$ and at the fundamental scale 
$\mu_{fundamental}$. The renormalisation group running of $g_t(\mu)$ varies 
logarithmically with the scale and, since the beta function $\beta_{g_t}$ 
is relatively small, the scale ratio $\mu_{fundamental}/\mu_{weak}$ required 
to generate the difference between $g_t(\mu_{fundamental})$ and 
$g_t(\mu_{weak})$ turns out to be 
the exponential of a large number. As well as predicting the value of 
$g_t(\mu_{fundamental})$ in terms of the fine structure constants, the MPP 
also predicts values (see Fig. \ref{fig:alphas}) for these fine structure 
constants $\alpha_i(\mu_{fundamental})$ when the SM is extended to the Family 
Replicated Gauge Group Model. Hence we have a partial understanding of why 
the SM beta functions are relatively small.

One crucial ingredient in the above MPP solution of the scale problem, and 
thus essentially of the hierarchy problem, was the existence of a very 
strongly bound state. This bound state of 6 top quarks and 6 anti-top 
quarks is then supposed to condense in a phase of the SM vacuum for 
which $\sVEV{\phi_{\sss\rm WS}} \sim \mu_{weak}$. As a crude estimate of 
the top quark Yukawa coupling at the phase transition between the vacua 
with and without such a condensate, we took the value that made the binding 
energy of this bound state just compensate for the sum, $12m_t$, of the 
constituent masses. So we suggest the existence of a bound state 
with a mass much less than $12m_t$, which could possibly be light 
enough to be produced at say the Tevatron or LHC. The problem 
with detecting such a particle is, however, that even were it 
energetically possible the production cross-section would be very low, 
if it was just crudely related to the cross section for producing 
six $t$-$\overline{t}$ pairs. Also, because of the strong binding, 
the usual top quark decays could easily be suppressed or forbidden for 
top quarks inside the bound state. Consequently a much smaller decay 
width would be expected for such a bound state than if it were only
loosely bound. Similarly the strong interactions of our speculated 
colour singlet bound state will be highly suppressed because of its very 
small size. It can only couple with transition dipole moments to 
other hadrons or to two gluons. The decay of the bound state into 
two gluons would require the simultaneous disappearance of 12 
quarks/antiquarks and be strongly suppressed. A simple interaction with 
two gluons could give the bound state particles a rather long range but 
weak interaction with hadrons. 

Concerning our solution of the scale problem and its relation to the 
hierarchy problem, we should remark that our Multiple Point Principle 
did not solve the technical hierarchy problem. By this remark we mean 
that, even if our scenario was perfectly correct, there would still be 
quadratic divergencies in the radiative corrections to the Higgs mass 
squared at each order of perturbation theory. However, in our picture, 
one should now renormalise or adjust the bare Higgs mass squared not 
simply to give the measured Higgs mass, or some other measured quantity, 
but rather to the requirement of the Multiple Point Principle. In 
other words, the bare Higgs mass squared and other bare parameters 
should be adjusted, by big contributions in each order of perturbation 
theory\footnote{The properties of the bound state and the phase 
containing its condensate can only be calculated in a 
non-perturbative way. So we would have to 
imagine a calculational procedure with partially non-perturbative 
effects included from the start.}, to enforce the equality of 
the energy densities of the three different vacua considered 
in section \ref{scale}. The renormalisation would now be to fit 
MPP rather than just experiment.

In our opinion, however, it is not so obvious that a solution to the 
hierarchy problem in the technical sense is really required, if 
we can solve the related ``scale problem''. The latter problem 
refers to the question: why is the electroweak scale so very small 
compared to the supposed ``fundamental'' scale taken to be the Planck 
scale or some unification scale? Indeed we claim to have solved 
this scale problem in our MPP scenario, even to the extent that we 
predicted the correct order of magnitude for the logarithm of the 
ratio of the two scales mentioned! That is to say, we claim to have 
solved the fine-tuning problem of the surprisingly low Higgs mass or 
Higgs field VEV compared to the ``fundamental scale''. However it 
should be admitted that our Multiple Point Principle has the nature 
of a fine-tuning postulate rather than a mechanism for explaining 
fine-tunings. We, so to speak, put the fine-tunings in a new and 
perhaps simpler form, leaving it as a next step to invent a mechanism 
for our unified form of fine-tuning postulate. Indeed we have already 
argued \cite{glasgowbrioni} that baby-universe like theories \cite{baby}, 
having a mild breaking of locality and causality, may contain the underlying 
physical explanation of the Multiple Point Principle.

As is well-known, one has to postulate an extremely small value for 
the cosmological constant (= energy density in the vacuum) for the 
vacuum in which we live today. It only requires a mild rephrasing 
of the postulate into the statement that ``many different vacua 
should have approximately zero energy density'', in order to unify 
the cosmological constant fine-tuning problem with our Multiple Point 
Principle\footnote{We thank L.~Susskind for this remark, relating 
the MPP to the cosmological constant problem.}. In this sense, we can 
claim to have unified the scale and cosmological constant problems.

However, the real success for this unification of two fine-tuning 
problems is that it also gives good predictions for the top quark 
and Higgs particle masses and even for the fine structure constants, 
when the SM is extended to the Family Replicated Gauge Group Model. 
Furthermore, as discussed in section \ref{masses}, this 
model provides a successful phenomenology for all the quark-lepton masses 
and mixing angles. So we conclude that the Multiple Point Principle 
together with the Family Replicated Gauge Group Model provide a 
viable understanding of the SM parameters. 

\section*{Acknowledgements}
We should like to congratulate Michael Danilov and his colleagues
on organising a very successful School. We should also like to
acknowledge helpful discussions with our collaborators Larisa
Laperashvili and Roman Nevzorov in Moscow.

\end{document}